\documentclass[aps,pre,preprint,superscriptaddress]{revtex4-1}
\usepackage[utf8]{inputenc}
\usepackage[T1]{fontenc}
\usepackage{mathptmx}
\usepackage{etoolbox}
\usepackage{amssymb}
\usepackage{color}
\usepackage{graphicx}
\usepackage{amsmath}
\usepackage{mathrsfs}
\usepackage{times}
\usepackage{subeqnarray}
\usepackage{cases}
\usepackage{dcolumn} 
\usepackage{import}
\begin{document}

\title{Eigenvector-based analysis of cluster synchronization in general complex networks of coupled chaotic oscillators}

\author{Huawei Fan}
\affiliation{School of Science, Xi'an University of Posts and Telecommunications, Xi'an 710121, China}
\affiliation{School of Physics and Information Technology, Shaanxi Normal University, Xi'an 710062, China}

\author{Ya Wang}
\affiliation{School of Physics and Information Technology, Shaanxi Normal University, Xi'an 710062, China}

\author{Xingang Wang}
\email[Email address: ]{wangxg@snnu.edu.cn}
\affiliation{School of Physics and Information Technology, Shaanxi Normal University, Xi'an 710062, China}

\begin{abstract}
Whereas topological symmetries have been recognized as crucially important to the exploration of synchronization patterns in complex networks of coupled dynamical oscillators, the identification of the symmetries in large-size complex networks remains as a challenge. Additionally, even though the topological symmetries of a complex network are known, it is still not clear how the system dynamics is transited among different synchronization patterns with respect to the coupling strength of the oscillators. We propose here the framework of eigenvector-based analysis to identify the synchronization patterns in the general complex networks and, incorporating the conventional method of eigenvalue analysis, investigate the emergence and transition of the cluster synchronization states. We are able to argue and demonstrate that, without a prior knowledge of the network symmetries, the method is able to predict not only all the cluster synchronization states observable in the network, but also the critical couplings where the states become stable and the sequence of these states in the process of synchronization transition. The efficacy and generality of the proposed method are verified by different network models of coupled chaotic oscillators, including artificial networks of perfect symmetries and empirical networks of non-perfect symmetries. The new framework paves a way to the investigation of synchronization patterns in large-size, general complex networks.
\end{abstract}
\date{\today }
\maketitle

\section{introduction}\label{intro}

As a universal phenomenon in complex systems of coupled dynamical units, synchronization has been broadly interested by researchers from different fields in the past decades~\cite{BOOK:Kuramoto,BOOK:ATW,BOOK:PRK,BOOK:SS}. Briefly, synchronization refers to the coherent motion of coupled oscillators, which is observed normally when the coupling strength between the oscillators exceeds a critical value. Depending on the correlation between the oscillators, synchronization can be realized in different forms, e.g., complete synchronization, phase synchronization and generalized synchronization~\cite{BOOK:PRK}. In synchronization studies, a key issue is to find, numerically or analytically, the critical coupling for generating synchronization. For systems consisting of linearly coupled identical chaotic oscillators, complete synchronization can be realized and the critical coupling for synchronization can be analyzed by the formalism of master stability function (MSF)~\cite{MSF-1,MSF-2,MSF-3}; for systems composed of nonlinearly coupled non-identical phase oscillators, the critical coupling characterizing the onset of phase synchronization can be estimated by some mean-field approaches~\cite{REV:JAA,OA:2008}. Whereas earlier studies of oscillator synchronization have been focusing on small-size systems of regular coupling structures~\cite{BOOK:KK}, e.g., the lattices and globally coupled systems, recent studies extend the studies to large-size systems with complex coupling structures, e.g., the synchronization in complex networks~\cite{SW:1998,BA:1999,NetBook:Newman,REV:SB2006,REV:Arenas,REV:LZH}. The adoption of complex networks opens up a new window to the study of oscillator synchronization, in which the important roles of network structure on synchronization have been revealed~\cite{NetSyn:WXF2002,NetSyn:Pecora2002,NetSyn:Nishikawa,NetSyn:Arenas}. 

For systems of coupled identical chaotic oscillators, an interesting phenomenon observed in modeling systems and experiments is that under certain circumstances the oscillators can be synchronized in groups, namely the phenomenon of cluster synchronization (CS)~\cite{CS:Hansel1993,CS:Hasler,CS:YZ,CS:AP,CS:ExpZY,CS:CRSW,CS:WXG2017,CS:MMN,CS:BiologyNet,ER:1999,CS:robot}. In CS, oscillators within each cluster are highly correlated, but not if the oscillators belong to different clusters. Compared to global synchronization, CS is normally observed at weaker couplings in the transition regime from desynchronization to global synchronization. The spatial distribution of the clusters on the network defines a pattern and, by varying the coupling strength, the same system can present different synchronization patterns. In the study of CS, two of the central questions are: (1) how to find the possible synchronization patterns for the given network structure and (2) how to estimate the critical couplings generating the patterns. For systems possessing regular coupling structures, e.g., the ring-structure network, the patterns are normally generated through the mechanism of symmetry breaking, and the spatial and dynamical properties of the patterns can be analyzed by methods such as eigenvalue-based analysis~\cite{Heagy:1995,Pecora:1998}. Yet challenges arise when the oscillators are coupled on complex networks ~\cite{CS:BAO,CS:OTT2007,SyncPattern,CS:WXG2014,Pecora2014,SynPat:Schaub,FS:2016,JDH:2019}. Different from the regular networks in which the synchronization patterns can be inspected visually, the patterns on complex networks are blurred by the network structures. As such, to identify and analyze the synchronization patterns on complex networks, special techniques and methods different from the conventional ones should be employed~\cite{ADM:2016}. 

One approach to analyzing the synchronization patterns in complex networks is exploiting the information of network topological symmetries~\cite{CS:BAO,CS:OTT2007,SyncPattern,CS:WXG2014,SynPat:Schaub,FS:2016,JDH:2019,Pecora2014,Recentadvances,LWJ-1,LWJ-2,NTTSCS,YC:2017,BC:2018,CS:WYF2019,CSWL:2020}. Briefly, if a set of oscillators whose permutations on the network do not change the network dynamics, the oscillators are said to be symmetric with each other and the set of symmetric oscillators have the potential to form a synchronization cluster~\cite{Recentadvances}. Whereas the symmetries of small-size networks can be figured out straightforwardly, the finding of all symmetries in large-size complex networks requires some sophisticated techniques~\cite{Pecora2014,NetworkSym:LYS2022,EEP:Cardoso,QuoNet,EEP:OClery}. Once the network symmetries are identified, the next step is to analyze the stability of the synchronization patterns associated with the symmetries. In doing this, a general approach is to decouple the dynamics of the CS state from the dynamics of the perturbations. This can be done by techniques such as irreducible representations (IRR) and generalized MSF~\cite{SyncPattern,CS:WXG2014,Pecora2014,YC:2017}, by which the critical couplings for generating the CS states can be estimated. Besides the approach of network-symmetry-based analysis, the CS states of complex networks can also be analyzed by methods such as external equitable partition (EEP) and simultaneous block diagonalization (SBD)~\cite{EEP:Cardoso,QuoNet,EEP:OClery,SBD:Irving,SBD:Zhang2020,SBD:Zhang2021,SBD:Panahi}. The EEP method concerns the dynamics of a small-size quotient network, and the contents of each cluster are defined according to the inputting signals~\cite{EEP:Cardoso,QuoNet,EEP:OClery}; the SBD method is based on partitions of the networked nodes, and is featured by a simultaneous block diagonalization of the network coupling matrix~\cite{SBD:Irving,SBD:Zhang2020,SBD:Zhang2021,SBD:Panahi}. 

The existing methods of CS analysis, however, encounter a major difficulty when dealing with realistic systems: perfect symmetry is hardly observed in real-world complex networks~\cite{Pecora2014,Recentadvances,SBD:Panahi}. In grouping network nodes into clusters, the existing methods require a perfect symmetry of the nodes~\cite{CS:BAO,CS:OTT2007,SyncPattern,CS:WXG2014,SynPat:Schaub,FS:2016,JDH:2019,Pecora2014,Recentadvances,LWJ-1,LWJ-2,NTTSCS,YC:2017,BC:2018,CS:WYF2019,CSWL:2020}, i.e., the exchanges of the symmetric nodes on the network does not affect the network dynamics. This rigorous requirement makes most of the clusters identified in typical complex networks trivial, in the sense that the clusters contain only several or just a single node~\cite{Golubitsky:1985}. Yet mesoscale synchronization clusters do exist in realistic systems, which are believed as playing crucial roles in realizing the system functions, e.g., the brain networks~\cite{CSinbN:Zhou1,CSinbN:Zhou2,CSinbN:Wang,CSCN:Huo}. In specific, empirical studies show that many realistic networks are composed of communities, with nodes inside each community being densely connected and the connections between communities are sparse~\cite{commnet:Newman}. When oscillators are coupled on a complex community network, distinct synchronization clusters can be observed and the contents of the clusters are consistent with the partition of the communities~\cite{CSinbN:Zhou1,commnet:Huang,commnet:Wang}. As both the inter- and intra-connections are randomly established, perfect symmetry does not exist in complex community networks in general, which, according to the existing methods of CS analysis, implies that distinct synchronization clusters can not be observed. The contradiction between theory and reality makes it necessary to generalize the current methods of CS analysis to the general complex networks, which is the major objective of our present work. Besides the concern of network symmetry, another question encountered in analyzing CS is the separated analysis of cluster partition and stability. To obtain the critical coupling associated with a specific CS state by the existing methods, one needs to first find the contents of the clusters associated with the state by some node-partition techniques, then construct the quotient network governing the dynamics of this CS state, and finally evaluate the stability of the CS state and obtain the critical coupling. As the quotient network is dependent on the partition of the clusters, the stability of the CS states needs to be evaluated individually and separately, making a complete analysis of the CS states in large-size complex networks time-consuming and inconvenient. The second objective of our present work is to combine the two-step analysis, namely cluster partition and stability analysis, into a one-step analysis, so as to facilitate the exploration of CS in large-size networks. 

The mission of our present work is to argue that the above objectives, i.e., extending the CS studies to general complex networks and combining the analysis of cluster identification and stability into a single analysis, can be accomplished within the framework of eigenvector-based analysis. To be more specific, we shall demonstrate in different network models that, without a priori knowledge of the network symmetries, the proposed method is able to predict not only the CS states to be emerged in synchronization transition, but also the range over which each CS state is observable in the parameter space and how the network dynamics is transited among the CS states as the coupling parameter is varying. The rest of the paper is organized as follows. In the following section, we shall present the theoretical framework used for analyzing CS in general complex networks. The applications of the proposed framework to different network models, including small-size networks of perfect symmetries and large-size complex networks of non-perfect symmetries, will be reported in Sec. III. Finally, discussions and conclusion will be given in Sec. IV.    

\section{Eigenvector-based analysis}

The dynamical system we consider here is a complex network of coupled identical chaotic oscillators, with the system dynamics described by the equations
\begin{equation}\label{model}
\dot{\mathbf{x}}_{i}=\mathbf{F}(\mathbf{x}_{i})+\varepsilon \sum^{N}_{j=1}w_{ij}\mathbf{H}(\mathbf{x}_{j}).
\end{equation}
Here, $i,j=1,\ldots,N$ are the node (oscillator) indices, $\mathbf{x}_{i}$ represents the state vector of the $i$th oscillator, $\mathbf{F}(\mathbf{x})$ denotes the dynamics of the isolated oscillators, $\mathbf{H}(\mathbf{x})$ defines the coupling function, and $\varepsilon$ is the uniform coupling strength. The coupling relationship of the oscillators is captured by the weighted matrix  $\mathbf{W}=\{w_{ij}\}$, with $w_{ij}>0$ denoting the strength of the coupling that node $i$ is received from node $j$. If nodes $i$ and $j$ are not connected, we set $w_{ij}=w_{ji}=0$. The diagonal elements of $\mathbf{W}$ are set as $w_{ii}=-\sum_{j(j\neq i)}w_{ij}$, i.e., $\mathbf{W}$ is a Laplacian matrix. The model described by Eq.~(\ref{model}) has been widely adopted in the literature for exploring the synchronization of coupled chaotic oscillators, which is a good approximation to the dynamics of many realistic systems in the vicinity of their functional states~\cite{BOOK:PRK,REV:SB2006,REV:Arenas}. 

As the oscillators are linearly coupled and $\mathbf{W}$ is a Laplacian matrix, the state of global synchronization is a solution to Eq.~(\ref{model}). Denote $\mathbf{s}$ as the manifold of global synchronization state, i.e., $\mathbf{s}=\mathbf{x}_{1}=\mathbf{x}_{2}=...=\mathbf{x}_{N}$, the starting point of our analysis is to evaluate the stability of the synchronization state in the presence of small, random perturbations. Let $\delta \mathbf{x}_{i}=\mathbf{x}_{i}-\mathbf{s}$ be infinitesimal perturbation added onto oscillator $i$ in the global synchronization state, the evolution of $\delta \mathbf{x}_{i}$ is governed by the variational equations
\begin{equation}\label{variational-eq}
\delta\dot{\mathbf{x}}_{i}=\mathbf{DF}(\mathbf{s})\delta\mathbf{x}_{i}+\varepsilon\sum^{N}_{j=1}w_{ij}\mathbf{DH}(\mathbf{s})\delta\mathbf{x}_{j},
\end{equation}
with $\mathbf{DF}(\mathbf{s})$ and $\mathbf{DH}(\mathbf{s})$ the Jacobian matrices of $\mathbf{F}$ and $\mathbf{H}$ evaluated on the synchronization manifold $\mathbf{s}$. Transforming Eq.~(\ref{variational-eq}) into the space spanned by the eigenvectors of $\mathbf{W}$, we have
\begin{equation}\label{decoupled-eq}
\delta\mathbf{\dot{y}}_{i}=[\mathbf{DF}(\mathbf{s})+\varepsilon\lambda_{ i}\mathbf{DH}(\mathbf{s})]\delta\mathbf{y}_{i}.
\end{equation}
Here, $\Delta\mathbf{Y}=[\delta \mathbf{y}_{1}, \delta \mathbf{y}_{2},..., \delta \mathbf{y}_{N}]^{T}=\mathbf{V}^{-1}[\delta \mathbf{x}_{1}, \delta \mathbf{x}_{2},..., \delta \mathbf{x}_{N}]^{T}$ are the perturbation modes in the new space, $\mathbf{V}$ is the transformation matrix composed by the eigenvectors of $\mathbf{W}$, and $0=\lambda_{1}>\lambda_{2}\geq\ldots\geq\lambda_{N}$ are the eigenvalues of $\mathbf{W}$. Among the $N$ modes, the mode associated with $\lambda_1$ describes the motion parallel to the synchronization manifold, and the modes associated with $\lambda_{2,\ldots,N}$ describe the motions transverse to the synchronization manifold. Following the MSF formalism~\cite{MSF-1,MSF-2,MSF-3}, we introduce the generic coupling strength $\sigma\equiv-\varepsilon\lambda$ and rewrite Eq.~(\ref{decoupled-eq}) as
\begin{equation}\label{msf}
\delta\mathbf{\dot{y}}_{i}=[\mathbf{DF}(\mathbf{s})-\sigma_i\mathbf{DH}(\mathbf{s})]\delta\mathbf{y}_{i}.
\end{equation}
For the global synchronization state to be stable, the necessary condition is that all the perturbation modes transverse to the synchronization state should be damping with time. That is, the largest conditional Lyapunov exponent, $\Lambda_i$, of Eq.~(\ref{msf}) should be negative for modes $i=2,\ldots,N$. As Eq.~(\ref{msf}) applies to all the perturbation modes, it is named the master equation of the perturbation dynamics~\cite{MSF-1}. By solving the master equation numerically (or analytically in some special cases), we can obtain the variation of $\Lambda$ with respect to $\sigma$, which defines the MSF curve. In the MSF curve, the regions with $\Lambda<0$ constitute the stable domain in the parameter space, and the regions with $\Lambda>0$ constitute the unstable domain. Depending on the nodal dynamics and the coupling function, the stable domain may have different forms, e.g., bounded or unbounded~\cite{MSF-3}. For simplicity, we consider here only the case of unbounded stable domain, in which $\Lambda$ becomes negative when $\sigma$ is larger than a critical value $\sigma_c$. In this case, global synchronization will be achieved when $\varepsilon>\varepsilon_c=\sigma_c/\lambda_2$, with $\lambda_2$ the 2nd largest eigenvalue of $\mathbf{W}$. So far, our analysis is identical to the standard MSF formalism~\cite{MSF-1,MSF-2,MSF-3}, in which the only role of the eigenvectors of the network coupling matrix $\mathbf{W}$ is to construct the transformation matrix $\mathbf{V}$, while the stability of the global synchronization state is dependent on only the eigenvalues of the network coupling matrix. In what follows, we are going to argue that the eigenvectors play an important role in exploring the CS states in general complex networks of coupled chaotic oscillators.

We move on to analyze the desynchronization state of the network when $\varepsilon<\varepsilon_c$. We consider first the situation when a single mode in the transverse space is unstable. As we are focusing on MSF of unbounded stable domain, the most unstable mode is associated with $\lambda_2$. That is, by decreasing $\varepsilon$ from $\varepsilon_c$, the mode $\delta\mathbf{y}_2$ is first emerged from the uniform background of global synchronization. Since the mode associated with $\lambda_1=0$ is always unstable, we thus have for this situation two unstable modes in fact, $\delta\mathbf{y}_1$ and $\delta\mathbf{y}_2$. Denote $\delta\mathbf{y}_1(t)$ and $\delta\mathbf{y}_2(t)$ as the modes of $\lambda_1$ and $\lambda_2$ at time $t$ respectively, we transfer them into the node space and obtain 
\begin{equation}\label{v2}
\delta\mathbf{x}_{i}(t)=v_{1,i}\delta\mathbf{y}_{1}(t)+v_{2,i}\delta\mathbf{y}_{2}(t).
\end{equation}
Here, $\mathbf{v}_1=\{v_{1,i}\}_{i=1,\ldots,N}$ is the eigenvector associated with $\lambda_1$, and $\mathbf{v}_2=\{v_{2,i}\}_{i=1,\ldots,N}$ is the eigenvector associated with $\lambda_2$. As $\lambda_1=0$, we have $v_{1,i}=1/\sqrt{N}$ for all the elements of $\mathbf{v}_1$. That is, the first term on the right-hand-side (RHS) of Eq.~(\ref{v2}) is identical for all the oscillators, and the perturbation that oscillator $i$ is away from the global synchronization state is solely determined by the second term on the RHS of the equation. More specifically, we have
\begin{equation}\label{v2-2}
\delta\mathbf{x}_{i}(t)=c(t)+v_{2,i}\delta\mathbf{y}_{2}(t),
\end{equation} 
with $c(t)$ a variable independent of the oscillator index. Therefore, by checking the values of $v_{2,i}$, the stability of the oscillators can be evaluated individually: the larger is $v_{2,i}$, the more unstable is oscillator $i$. In particular, if $v_{2,i}=v_{2,j}$, the states of the pair of oscillators $i$ and $j$ will be identical during the process of system evolution, i.e., they are completely synchronized. The set of oscillators with the identical eigenvector element thus forms a synchronization cluster. That is, by checking just the elements of the eigenvector $\mathbf{v}_2$, we are able to identify all the synchronization clusters when the mode of $\lambda_2$ is unstable. 

We consider next the situation when several perturbation modes are unstable. This occurs when $\varepsilon$ is much less than $\varepsilon_c$ and means that the network is deeply desynchronized. We assume that the statistical properties of the oscillators, e.g., the measure of the oscillator trajectory, can be approximated by that of the isolated oscillator. (The validity of this approximation will be verified by numerical simulations later.) With this approximation, the stability of the perturbation modes is still governed by Eq.~(\ref{msf}). We assume further that $\tilde{m}$ modes are unstable, and the eigenvectors of the unstable modes are $\mathbf{v}_k$, with $k=2,\ldots,\tilde{m}+1$. Note that for the case of unbounded MSF curve, the perturbation modes are destabilized in sequence by the descending order of the eigenvalues, with the mode of $\lambda_2$ being the first one and the mode of $\lambda_N$ being the last one. Specifically, the $k$th perturbation mode becomes unstable when $\varepsilon<\varepsilon_k=-\sigma_k/\lambda_k$. When $\tilde{m}$ modes are unstable in the network, the perturbation of the $i$th oscillator can be written as
\begin{equation}\label{vk}
\delta\mathbf{x}_{i}(t)=c(t)+\sum_{k=2}^{\tilde{m}+1}v_{k,i}\delta\mathbf{y}_{k}(t).
\end{equation}
Still, $c(t)$ represents the perturbation in parallel to the synchronization manifold, which is independent of the oscillator index. Now, for oscillators $i$ and $j$ to be completely synchronized, the necessary condition is that $v_{k,i}=v_{k,j}$ for all the unstable modes $k=2,\ldots,\tilde{m}+1$. Hence, by checking only the eigenvectors of the $\tilde{m}$ unstable modes, we are able to find all the synchronization clusters on the desynchronized network.

The above analysis applies to only networks of perfect symmetries. In specific, for two oscillators to be synchronized, their elements in the eigenvectors of all the unstable modes should be exactly the same. From the point of view of network topology, this means that the two oscillators are identical and indistinguishable. This restricts seriously the application of the above analysis to the general complex networks, in which links are normally weighted and perfect symmetry does not exist in general. To cope with this problem, we loosen the requirement for cluster synchronization, and regard oscillators $i$ and $j$ as synchronized if the time-averaged error $\delta \tilde{x}_{i,j}=\left<|\mathbf{x}_i-\mathbf{x}_j|\right>$ between them is smaller to a small threshold. To investigate theoretically the CS behaviors in the general complex networks, now the question becomes: for a moderate coupling strength by which global network synchronization is impossible, can we predict the contents of the loosely defined synchronization clusters based on the information of the network coupling matrix and, furthermore, the ranges over which the synchronization clusters are observed in the parameter space of the uniform coupling strength? Noticing that $\mathbf{x}_i-\mathbf{x}_j=\delta\mathbf{x}_i-\delta\mathbf{x}_j$, we have from Eq.~(\ref{vk}) the relation
\begin{equation}\label{error1}
\delta \tilde{x}_{i,j}=\left<\left |\sum_{k=2}^{\tilde{m}+1}(v_{k,i}-v_{k,j})\delta\mathbf{y}_{k}(t)\right|\right>.
\end{equation}
For a deeply desynchronized network, the amplitudes of the unstable modes are approximately the same, i.e., $|\delta\mathbf{\hat{y}}(t)|\approx|\delta\mathbf{y}_k(t)|$. With this approximation, Eq.~(\ref{error1}) can be simplified as
\begin{equation}\label{error}
\delta \tilde{x}_{i,j}=\left<\left | \delta \tilde{e}_{i,j} \delta\mathbf{\hat{y}}(t)\right|\right>,
\end{equation}
with 
\begin{equation}\label{eigendis}
\delta \tilde{e}_{i,j}=\sum_{k=2}^{\tilde{m}+1}|v_{k,i}-v_{k,j}|
\end{equation}
the distance between nodes $i$ and $j$ in the eigenvector space. To facilitate the analysis of CS in general complex networks (so that the thresholds used for finding the clusters are independent of the coupling strength), here we adopt the normalized distance $\delta e_{i,j}=\delta \tilde{e}_{i,j}/\tilde{m}$ to characterize the topological difference between nodes $i$ and $j$. As $\delta\mathbf{\hat{y}}$ is independent of the node index, we finally have 
\begin{equation}\label{theory}
\delta \tilde{x}_{i,j} \propto \delta e_{i,j}.
\end{equation}
The Eq.~(\ref{theory}) is our main theoretical result, which tells how the synchronization degree of two oscillators (characterized by $\delta\tilde{x}_{i,j}$) can be inferred from the network topology (characterized by $\delta e_{i,j}$). We note that both $\delta\tilde{x}_{i,j}$ and $\delta e_{i,j}$ are dependent on the coupling strength, but the relation described by Eq.~(\ref{theory}) holds for any coupling strength.

In applications, the above framework of eigenvector-based analysis is implemented as follows. The first step is to cacluate the eigenvalues $\{\lambda_i\}_{i=1,\ldots,N}$ and the corresponding eigenvectors $\{\mathbf{v}_i\}_{i=1,\ldots,N}$ of the network coupling matrix $\mathbf{W}$. The second step is to find the boundary of the stable regime of the MSF curve, i.e., the value of $\sigma_c$, which can be obtained by solving Eq.~(\ref{msf}) numerically. The third step is to find the set of unstable modes, $\{\lambda_k\}_{k=2,\ldots,\tilde{m}+1}$, for a specific coupling strength of interest, and then calculate the matrix of eigenvector distance, $\{\delta e_{i,j}\}_{\tilde{m}\times \tilde{m}}$, according to Eq.~(\ref{theory}). Finally, we find from the distance matrix the set of oscillators forming a cluster. For complex networks of perfect symmetries, the contents of each cluster are identified by the requirement $\delta e_{i,j}=0$ (i.e., the eigenvector distance between the oscillators within each cluster is $0$); while for the general complex networks, the contents of each cluster are identified by requiring $\delta e_{i,j}$ to be smaller than a predefined threshold.  

\section{Applications}

We next apply the theoretical framework to investigate the CS behaviors in different network models, including small-size networks of perfect symmetries, large-size complex networks of community structures, and two empirical neural networks.

\subsection{Small-size networks}

\begin{figure*}[tbp]
\begin{center}
\includegraphics[width=0.87\linewidth]{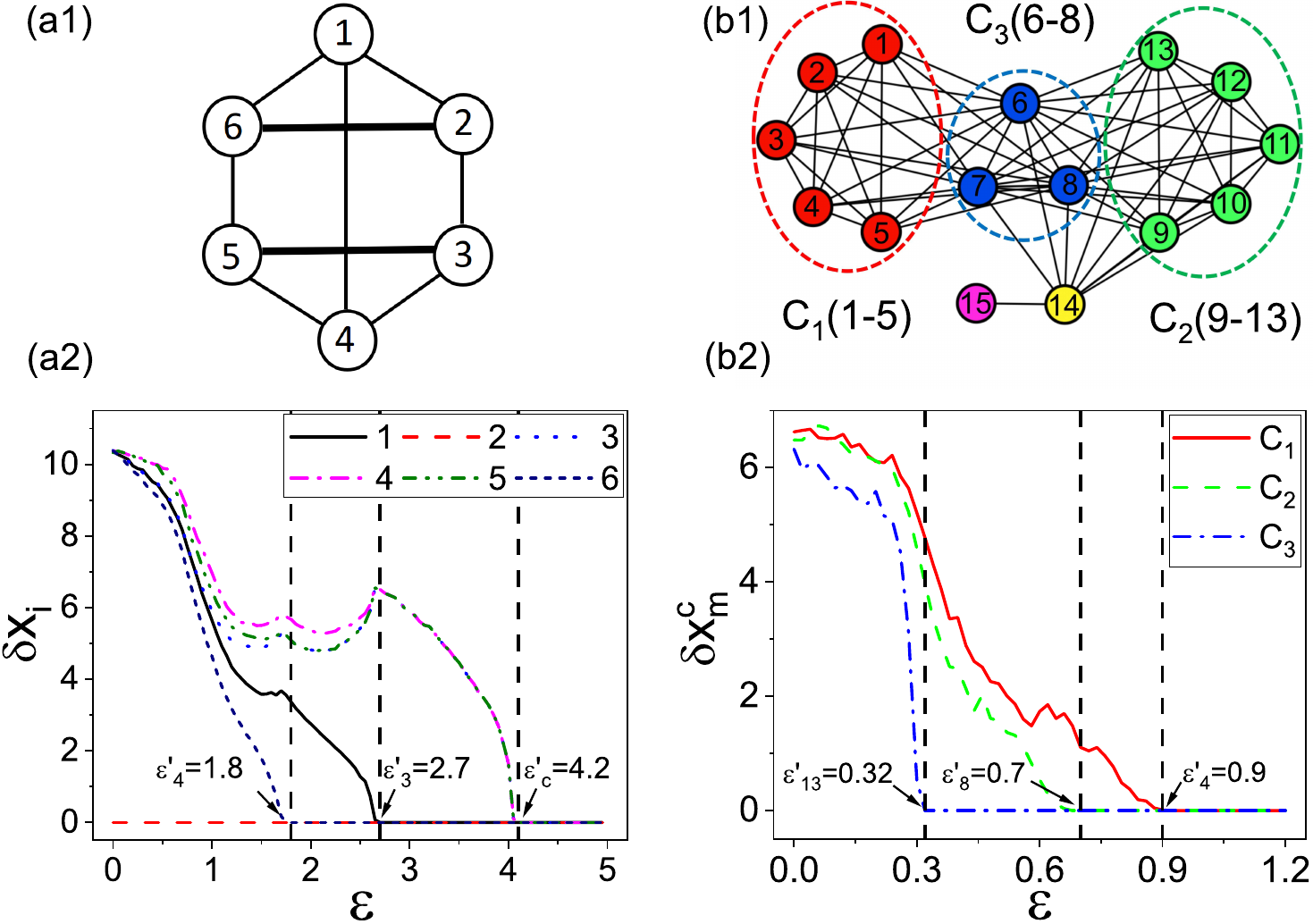}
\caption{CS in small-size networks of perfect symmetries. (a1) The structure of the $6$-node network. The network dynamics is described by Eq.~(\ref{N6net}). (a2) By model simulations, the CS states observed in synchronization transition. $\delta x_{i}=\langle |x_{i}-x_{2}|\rangle_{T}$ is the synchronization error between oscillator $i$ and the $2$nd oscillator. $\varepsilon$ denotes the uniform coupling strength. In the range of $\varepsilon\in(1.8,2.7)$, two synchronization pairs, $(2,6)$ and $(3,5)$, are generated. In the range of $\varepsilon\in(2.7,4.2)$, two synchronization clusters, $(1,2,6)$ and $(3,4,5)$, are formed. Global network synchronization is achieved when $\varepsilon>\varepsilon'_c\approx 4.2$. (b1) The network structure of the Nepal power grid. The network nodes are partitioned into three non-trivial clusters, $\mathbf{C}_1=\{1,\ldots,5\}$, $\mathbf{C}_2=\{9,\ldots,13\}$ and $\mathbf{C}_3=\{6,7,8\}$, and two trivial clusters, $\mathbf{C}_4=\{14\}$ and $\mathbf{C}_5=\{15\}$. (b2) By numerical simulations, the variation of the synchronization errors of the non-trivial clusters, $\delta x^c_m=\sum_{i,j\in C_{m}}\langle|x_{i}-x_{j}|\rangle_{T}/{n_{m}(n_{m}-1)}$ (with $m=1,2,3$ the cluster index and $n_m$ the number of nodes in cluster $m$), with respect to $\varepsilon$. The 1st, 2nd and 3rd clusters are synchronized at about $\varepsilon=0.9$, $0.7$ and $0.32$, respectively.}
\label{fig1}
\end{center}
\end{figure*}

We start with a toy network model of perfect symmetries. The network structure is shown in Fig.~\ref{fig1}(a1), which contains $N=6$ nodes and $9$ links. Following Ref.~\cite{CS:WXG2014}, we set the weight of the link between nodes $2$ and $6$ as $1.5$, and the same weight is arranged for the link between nodes $3$ and $5$. The weights of the other links are all set as unity. The coupling matrix of the network reads
\begin{equation}
\mathbf{W}=\left(
\begin{array}{cccccc}
-3 & 1 & 0 & 1 & 0 & 1 \\
 1 &-3.5 & 1 & 0 & 0 & 1.5 \\
 0 & 1 &-3.5 & 1 & 1.5 & 0\\
 1 & 0 & 1 &-3 & 1 & 0\\
 0 & 0 & 1.5 & 1 & -3.5 & 1 \\
 1 & 1.5 & 0 & 0 & 1 & -3.5 \\
\end{array}
\right).
\end{equation}
The eigenvalues of $\mathbf{W}$ are
\begin{equation}
(\lambda_1,\lambda_2,\lambda_3,\lambda_4,\lambda_5,\lambda_6)=(0, -2, -3, -4, -5, -6),
\end{equation}
and the eigenvector matrix is 
\begin{equation}\label{evt}
\mathbf{V}=(\mathbf{v}_1,\mathbf{v}_2,\mathbf{v}_3,\mathbf{v}_4,\mathbf{v}_5,\mathbf{v}_6)=\left(
\begin{array}{cccccc}
0.41&-0.41 &0.58 & 0  &0.58  & 0   \\
0.41&-0.41 &-0.29&0.5 &-0.29 & 0.5 \\
0.41& 0.41 &-0.29&0.5 & 0.29 &-0.5 \\
0.41& 0.41 &0.58 & 0  &-0.58 & 0   \\
0.41& 0.41 &-0.29&-0.5& 0.29 & 0.5 \\
0.41&-0.41 &-0.29&-0.5&-0.29 &-0.5 \\
\end{array}
\right).
\end{equation}

We adopt the chaotic Lorenz oscillator to describe the nodal dyamimcs~\cite{Lorenz}. The dynamics of the network reads
\begin{equation}
\begin{cases}
\dot{x_i}=\alpha(y_i-x_i), \\
\dot{y_i}=x_i(r-z_i)-y_i+\varepsilon\sum w_{ij}x_j,\\
\dot{z_i}=x_i y_i-bz_i.
\end{cases}
\label{N6net}
\end{equation}
Here, $i,j =1,\ldots,N$ are the oscillator indices, and the coupling function is chosen as $\mathbf{H}(\mathbf{x})=[0,x,0]^{T}$, i.e., the $x$ variable is coupled to the $y$ variable. The parameters of the oscillators are all set as $(\alpha,r,b)=(10,35,8/3)$, with which the oscillators present chaotic motion (the largest Lyapunov exponent is about $1.05$). To characterize the synchronization relationship among the oscillators, we set oscillator $2$ as the reference oscillator and calculate the synchronization error $\delta x_{i}=\langle |x_{i}-x_{2}|\rangle_{T}$, where $\langle\cdot\rangle_{T}$ denotes the time average function. In simulations, the system dynamics is evolved numerically by the fourth-order Runge-Kutta method with the time step $\delta t=1\times 10^{-2}$. In calculating $\delta x_i$, the system is firstly evolved for a transient period of $T_{0}=1\times10^{4}$ so as to remove the impact of the initial conditions, and the results are averaged over a period of $T=1\times10^{4}$.

Following the MSF formalism, we first calculate numerically the variation of the largest Lyapunov exponent, $\Lambda$, of the master equation [i.e. Eq.(\ref{msf})] with respect to the generic coupling strength, $\sigma$. The results show that $\Lambda<0$ when $\sigma>\sigma_c\approx 8.3$. The critical couplings where the modes of $\lambda_2$, $\lambda_3$, $\lambda_4$, $\lambda_5$ and $\lambda_6$ become unstable thus are $\varepsilon_2=\varepsilon_c\approx 4.2$, $\varepsilon_3\approx 2.8$, $\varepsilon_4\approx 2.1$, $\varepsilon_5\approx 1.7$ and $\varepsilon_6\approx 1.4$, respectively. By the framework of eigenvector-based analysis, we next predict the possible CS states to be observed in the process of network desynchronization, starting from the global synchronization state generated at a larger coupling strength $\varepsilon>\varepsilon_c=\varepsilon_2$. When $\varepsilon$ crosses $\varepsilon_2$ from above, the mode of $\lambda_2$ will be unstable. According to the analysis presented in Sec. II, the specific form of the CS state to be generated from the uniform background of global synchronization is determined by the eigenvector $\mathbf{v}_2$. As shown in Eq.~(\ref{evt}) (the 2nd column of the eigenvector matrix), we have $v_{2,1}=v_{2,2}=v_{2,6}=-0.41$ and $v_{2,3}=v_{2,4}=v_{2,5}=0.41$. The eigenvector distances therefore are $\delta e_{i,j}=0$ for $i,j\in\{1,2,6\}$ or $i,j\in \{3,4,5\}$. That is, the theory predicts that when only the mode of $\lambda_2$ is destabilized, the network nodes are synchronized into two clusters, $\mathbf{C}_1=\{1,2,6\}$ and $\mathbf{C}_2=\{3,4,5\}$. Decreasing further $\varepsilon$, the mode of $\lambda_3$ will be destabilized when $\varepsilon$ crosses $\varepsilon_3$. By checking the eigenvector $\mathbf{v}_3$ Eq.~(\ref{evt}) (the 3rd column of the eigenvector matrix), we see that $v_{3,1}=v_{3,4}=0.58$, $v_{3,2}=v_{3,3}=v_{3,5}=v_{3,6}=-0.29$. As both the modes of $\lambda_2$ and $\lambda_3$ are unstable [i.e., $\tilde{m}=2$ in Eq.~(\ref{eigendis})], the eigenvector distances between the nodes thus are $\delta e_{2,6}=0$, $\delta e_{3,5}=0$, and $\delta e_{i,j}>0$ for other node pairs. Hence, the theory predicts that in this case the network contains two synchronization pairs, $\mathbf{C}_1=\{2,6\}$ and $\mathbf{C}_2=\{3,5\}$, and two desynchronized nodes, $\mathbf{C}_3=\{1\}$ and $\mathbf{C}_4=\{4\}$. Finally, as $\varepsilon$ falls below $\varepsilon_4$, the mode associated with $\lambda_4$ will be unstable. As $v_{4,1}=v_{4,4}=0$, $v_{4,2}=v_{4,3}=0.5$ and $v_{4,5}=v_{4,6}=-0.5$ (see the 4th in the eigenvector matrix), the eigenvector distance between any pair of nodes is larger than $0$. That is, the theory predicts that when the mode of $\lambda_4$ is unstable ($\varepsilon<\varepsilon_4$), no synchronization is observed between the oscillators, and the network is completely desynchronized.  

The above predictions are well verified by numerical simulations. Shown in Fig.~\ref{fig1}(a2) are the variation of the synchronization errors, $\delta x_{i}=\langle |x_{i}-x_{2}|\rangle_{T}$, with respect to the uniform coupling strength, $\varepsilon$. We see that the network is globally synchronized when $\varepsilon>\varepsilon'_c\approx 4.2$. In the range of $\varepsilon\in (\varepsilon'_3\approx 2.7,\varepsilon'_c)$, the oscillators are organized into two synchronizastion clusters, $\mathbf{C}_1=\{1,2,6\}$ and $\mathbf{C}_2=\{3,4,5\}$. In the range of $\varepsilon\in (\varepsilon'_4\approx 1.8,\varepsilon'_3)$, the network contains two synchronization pairs, $(2,6)$ and $(3,5)$, and two desynchronized nodes (oscillators $1$ and $4$). Finally, when $\varepsilon<\varepsilon'_4$, no synchronization is observed between the oscillators, and the network is completely desynchronized. The critical couplings of the CS states, as well as the order of their emergences in the process of network desynchronization, are in good agreement with the theoretical predictions.

The second network model of perfect symmetries employed in our studies is the Nepal power-grid~\cite{Pecora2014}. The network structure of the Nepal power-grid is plotted in Fig.~\ref{fig1}(b1), which consists of $N=15$ nodes and $L=62$ unweighted, non-directed links. Still, we describe the nodal dynamics by the chaotic Lorenz oscillator and couple the oscillators through the function $\mathbf{H}(\mathbf{x})=[0,x,0]^{T}$. The eigenvalues of the network coupling matrix are 
\begin{equation}
\small{
\{\lambda_i\}_{i=1,\ldots,15}=(0, -0.94, -3.10, -8, -8, -8, -8, -9, -9, -9, -9, -9.94, -14, -14, -14.03),
}
\end{equation}
and the eigenvector matrix, $\mathbf{V}=(\mathbf{v}_1,\ldots,\mathbf{v}_{15})$, reads
\begin{equation}\label{pgevector}
\footnotesize
\left(
\begin{array}{ccccccccccccccc}
0.26&-0.11 &-0.32& 0.71&-0.12& 0   &-0.53& 0   & 0   & 0   & 0   & 0.01& 0   & 0   & 0.14\\
0.26&-0.11 &-0.32&-0.71&-0.12& 0   &-0.53& 0   & 0   & 0   & 0   & 0.01& 0   & 0   & 0.14\\
0.26&-0.11 &-0.32& 0   & 0.48& 0.71& 0.27& 0   & 0   & 0   & 0   & 0.01& 0   & 0   & 0.14\\
0.26&-0.11 &-0.32& 0   & 0.48&-0.71& 0.27& 0   & 0   & 0   & 0   & 0.01& 0   & 0   & 0.14\\
0.26&-0.11 &-0.32& 0   &-0.72& 0   & 0.53& 0   & 0   & 0   & 0   & 0.01& 0   & 0   & 0.14\\
0.26&-0.07 & 0.01& 0   & 0   & 0   & 0   & 0   & 0   & 0   & 0   &-0.03& 0.72&-0.39&-0.51\\
0.26&-0.07 & 0.01& 0   & 0   & 0   & 0   & 0   & 0   & 0   & 0   &-0.03&-0.02& 0.82&-0.51\\
0.26&-0.07 & 0.01& 0   & 0   & 0   & 0   & 0   & 0   & 0   & 0   &-0.03&-0.70&-0.43&-0.51\\
0.26&-0.05 & 0.29& 0   & 0   & 0   & 0   & 0.74& 0.21& 0.43& 0.14& 0.17& 0   & 0   & 0.14\\
0.26&-0.05 & 0.29& 0   & 0   & 0   & 0   &-0.15&-0.19&-0.38& 0.77& 0.17& 0   & 0   & 0.14\\
0.26&-0.05 & 0.29& 0   & 0   & 0   & 0   & 0.20& 0.10&-0.69&-0.53& 0.17& 0   & 0   & 0.14\\
0.26&-0.05 & 0.29& 0   & 0   & 0   & 0   &-0.20&-0.74&-0.36&-0.30& 0.17& 0   & 0   & 0.14\\
0.26&-0.05 & 0.29& 0   & 0   & 0   & 0   &-0.59& 0.61& 0.28&-0.09& 0.17& 0   & 0   & 0.14\\
0.26& 0.06 & 0.23& 0   & 0   & 0   & 0   & 0   & 0   & 0   & 0   &-0.92& 0   & 0   & 0.17\\
0.26& 0.95 &-0.11& 0   & 0   & 0   & 0   & 0   & 0   & 0   & 0   & 0.10& 0   & 0   &-0.01\\
\end{array}
\right).
\end{equation}
While CS in the Nepal power-grid network has been studied in the literature~\cite{Pecora2014,LWJ-1}, the existing studies rely on a priori knowledge of the network symmetries. In what follows, we are going to show that, without knowing the network symmetries, the CS states can be well analyzed by the framework of eigenvector-based analysis.

We now analyze the CS states emerged in the process of desynchronization transition for the Nepal power-grid network, starting from the global synchronization state. The critical coupling for the whole network to be synchronized is $\varepsilon_c=\varepsilon_2=-\sigma_c/\lambda_2\approx 8.8$. When $\varepsilon$ crosses $\varepsilon_c$ from above, the mode of $\lambda_2$ will be unstable and, according to the method of eigenvector analysis, the synchronization clusters can be identified from the elements of $\mathbf{v}_2$ (the 2nd column of the matrix $\mathbf{V}$). As $v_{2,1}=\ldots=v_{2,5}=-0.11$, $v_{2,6}=v_{2,7}=v_{2,8}=-0.07$ and $v_{2,9}=\ldots=v_{2,13}=-0.05$, we have three clusters in this case, $\mathbf{C}_1=(1,\ldots,5)$, $\mathbf{C}_2=(9,\ldots,13)$ and $\mathbf{C}_3=(6,7,8)$, with the eigenvector distance being $0$ for nodes within the same cluster. The contents of the clusters are kept unchanged when the mode of $\lambda_3$ becomes unstable, as the partition of the elements of $\mathbf{v}_3$ is identical to that of $\mathbf{v}_2$. However, when the mode of $\lambda_4$ is destabilized ($\varepsilon<\varepsilon_4\approx 1.04$), the elements of $\mathbf{v}_4$ (the 4th column of the matrix $\mathbf{V}$) suggest that the 1st cluster will be broken, while the 2nd and 3rd clusters hold still. That is, the theory predicts that in the range of $\varepsilon\in (\varepsilon_4,\varepsilon_c)$, three synchronization clusters will be observed on the network, with the contents of the 1st, 2nd and 3rd clusters being defined by $\mathbf{C}_1$, $\mathbf{C}_2$ and $\mathbf{C}_3$, respectively. By decreasing $\varepsilon$ further, the eigenvector distances of the nodes are kept unchanged till the critical coupling $\varepsilon_8=0.92$ is met, where the mode of $\lambda_8$ becomes unstable. By checking the elements of $\mathbf{v}_8$ in Eq.~(\ref{pgevector}) (the 8th column), we see that elements of the nodes in cluster $2$ are non-identical, suggesting the breaking of the 2nd cluster at $\varepsilon_8$. As the 1st cluster has already been broken at $\varepsilon_4$, only cluster 3 survives in this case. Finally, at the critical coupling $\varepsilon_{13}\approx 0.6$, the mode of $\lambda_{13}$ is destabilized and, according to the elements of $\mathbf{v}_{13}$, the eigenvector distances between nodes in the 3rd cluster will be all larger than $0$. As such, the 3rd cluster will be desynchronized at $\varepsilon_{13}$. Summarizing up the results, the theory predicts that: (1) the network contains three synchronization clusters, $\mathbf{C}_{1,2,3}$, when $\varepsilon\in (\varepsilon_4,\varepsilon_c)$; (2) two synchronization clusters, $\mathbf{C}_{2,3}$, are formed when $\varepsilon\in (\varepsilon_8,\varepsilon_4)$; (3) a single synchronization cluster, $\mathbf{C}_3$, is left when $\varepsilon\in (\varepsilon_{13},\varepsilon_8)$; and (4) no synchronization cluster is observed when $\varepsilon<\varepsilon_{13}$.   

By the approach of numerical simulations, we plot in Fig.~\ref{fig1}(b2) the variation of the cluster-based synchronization error, $\delta x^c$, with respect to the coupling strength, $\varepsilon$. Here, cluster-based synchronization error is defined as $\delta x^c_m=\sum_{i,j\in C_{m}}\langle|x_{i}-x_{j}|\rangle_{T}/{n_{m}(n_{m}-1)}$, with $m=1,2,3$ the cluster index and $n_m$ the number of nodes in cluster $m$. Clearly, cluster $m$ is synchronized when $\delta x^c_m=0$. Figure~\ref{fig1}(b2) shows that: (1) three synchronization clusters are generated when $\varepsilon>\varepsilon'_4\approx 0.9$, with the contents of the clusters identical to the ones predicted by the theory; (2) clusters $2$ and $3$ are synchronized when $\varepsilon\in (\varepsilon'_8\approx 0.7, \varepsilon'_4)$; (3) only cluster $3$ is synchronized when $\varepsilon\in (\varepsilon'_{13}\approx 0.32, \varepsilon'_8)$; and (4) no cluster is synchronized when $\varepsilon<\varepsilon'_{13}$. We see that, though the framework of eigenvector-based analysis can not predict precisely the critical couplings where the clusters become unstable, it does predict accurately the patterns of the CS states and, additionally, the sequence of these states in the process of network desynchronization.

\subsection{Complex community networks}

So far the framework of eigenvector-based analysis has been utilized to investigate the CS states in only small-size networks of perfect symmetries. It remains as not clear whether the framework can be applied to general complex networks in which perfect symmetries do not exist. We next check the efficacy of the proposed framework in analyzing the CS behaviors of complex community networks -- the type of network that is representative to many realistic systems~\cite{commnet:Newman}. In constructing the complex community network, we first divide the $N$ nodes into $M$ communities of equal size, and then connect nodes within each community by the probability $p_1$ and nodes from different communities by the probability $p_2$, with $p_2<p_1$. To demonstrate the generality of the theoretical framework, here we adopt the chaotic Hindmarsh-Rose (HR) oscillator as the nodal dynamics~\cite{HR:1984}. The dynamics of isolated HR oscillators is governed by equations $[\dot{x},\dot{y},\dot{z}]=[y-x^{3}+3x^{2}-z+I, 1-5x^{2}-y, r[s(x+1.6)-z]]^{T}$. The parameters of the oscillators are set as $(r,s,I)=(6\times 10^{-3}, 4, 3.2)$, by which the oscillators present the chaotic motion (the largest Lyapunov exponent is about $1.3\times 10^{-2}$). The oscillators are coupled by gap junctions, with the coupling function being described as $\mathbf{H}(\mathbf{x})=[x,0,0]^{T}$. To make the network dynamic stable, here we adopt the strategy of normalized couplings~\cite{Coupling-1,Coupling-2}, i.e., $w_{ij}=a_{ij}/k_{i}$, with $\{a_{ij}\}_{N\times N}$ the binary adjacency matrix capturing the network structure and $k_i=\sum a_{ij}$ the degree node $i$. The diagonal elements of $\mathbf{W}$ are all set as $-1$. Though the network links are weighted and directed, the eigenvalues of the coupling matrix are all real~\cite{Coupling-1,Coupling-2}.

\begin{figure}[tbp]
\begin{center}
\includegraphics[width=0.65\linewidth]{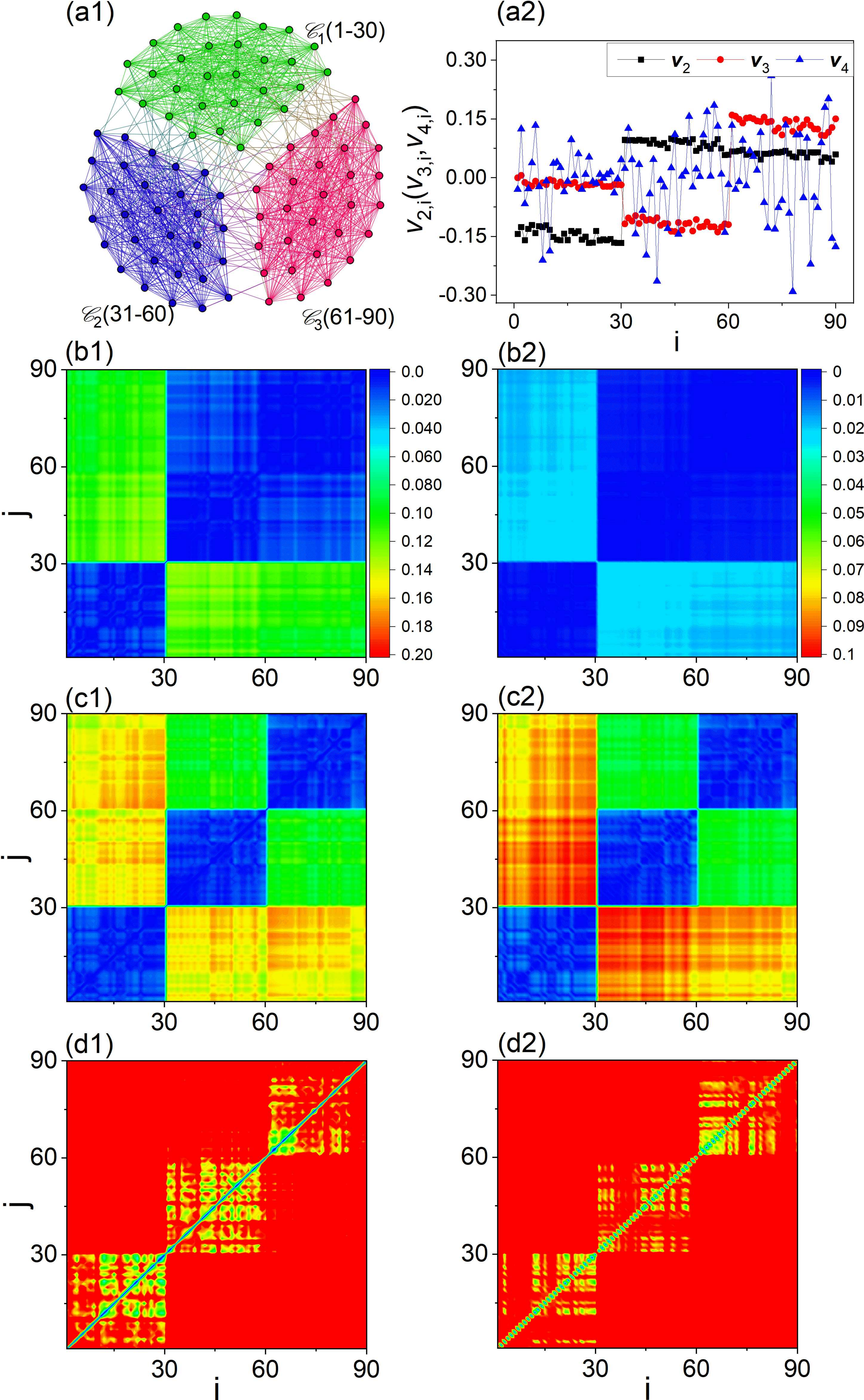}
\caption{CS in complex community networks of non-perfect symmetry. (a1) The structure of the community network, which contains $N=90$ nodes and $M=3$ communities of equal size. (a2) The elements of the eigenvectors $\mathbf{v}_2$, $\mathbf{v}_3$ and $\mathbf{v}_4$. (b1) The matrix of eigenvector distance, $\{\delta e_{i,j}\}_{i,j=1,\ldots,N}$, when the mode of $\lambda_2$ is unstable. (c1) The matrix of eigenvector distance when the modes of $\lambda_2$ and $\lambda_3$ are unstable. (d1) The matrix of eigenvector distance when the modes of $\lambda_2$, $\lambda_3$ and $\lambda_4$ are unstable. (b2) The matrix of synchronization error, $\{\delta x_{i,j}\}_{i,j=1,\ldots,N}$, for $\varepsilon=7.6$. (c2) The matrix of synchronization error for $\varepsilon=6$. (d2) The matrix of synchronization error for $\varepsilon=0.5$.}
\label{fig2}
\end{center}
\end{figure}

We start by analyzing the CS behaviors in a trivial complex community network consisting of $N=90$ nodes and $M=3$ communities. To distinct the community feature, here we set the connecting probabilities as $p_1=0.9$ (for the intra-community links) and $p_2=0.1$ (for the inter-community links). The network structure is plotted in Fig.~\ref{fig2}(a1), in which nodes are ordered according to the communities, $\mathcal{C}_1=(1,\dots,30)$, $\mathcal{C}_2=(31,\dots,60)$ and $\mathcal{C}_3=(61,\dots,90)$. As the links are randomly added among the nodes, there is no perfect symmetry in the network structure, which has been confirmed by the algorithm developed from computational group theory~\cite{Pecora2014}. While previous studies have shown that the synchronization clusters in community networks can be inferred from the eigenvector associated with the leading mode~\cite{commnet:Huang,commnet:Wang}, we are going to demonstrate in the following that the framework of eigenvector-based analysis is capable of offering more information about the CS behaviors. In applying the theoretical framework, we first calculate from the network coupling matrix the eigenvalues and the associated eigenvectors. The three largest non-trivial eigenvalues are $\lambda_{2}=-0.12$, $\lambda_{3}=-0.14$ and $\lambda_{4}=-0.88$. By simulating the master equation of the HR oscillator [see Eq.~(\ref{msf})], we have $\Lambda<0$ for $\sigma>\sigma_c\approx 0.94$. The critical coupling for the modes of $\lambda_2$, $\lambda_3$ and $\lambda_4$ therefore are $\varepsilon_2\approx 7.8$ (which is also the critical coupling for global synchronization), $\varepsilon_3\approx 6.7$ and $\varepsilon_4\approx 1.1$. Shown in Fig.~\ref{fig2}(a2) are the elements of the eigenvectors $\mathbf{v}_2$, $\mathbf{v}_3$ and $\mathbf{v}_4$. We see that the elements of $\mathbf{v}_2$ are clearly seperated into two groups, $v_{2,i}\approx -0.15$ for $i=1,\ldots,30$ and $v_{2,j}\approx 0.15$ for $j=31,\ldots,90$. In terms of the eigenvector distance defined by Eq.~(\ref{eigendis}), this means that $\delta e_{ij}\approx 0$ only for $i,j\in \mathcal{C}_1$ or $i,j\in \mathcal{C}_2\cup\mathcal{C}_3$. To have a global picture on the distribution of eigenvector distances, we plot in Fig.~\ref{fig2}(b1) the matrix $\{\delta e_{ij}\}_{i,j=1,\ldots,N}$, for the coupling strength $\varepsilon=7.6$ (by which the mode of $\lambda_2$ is unstable). Clearly, the oscillators are divided into two groups. According to the method of eigenvector analysis, we thus predict that as $\varepsilon$ decreases and crosses $\varepsilon_2$, the oscillators are synchronized into two clusters, $\mathbf{C}_1=(1,\dots,30)$, $\mathbf{C}_2=(31,\ldots,90)$. Decreasing $\varepsilon$ to $\varepsilon_3$, the mode of $\lambda_3$ will be unstable. By checking the elements of $\mathbf{v}_3$ in Fig.~\ref{fig2}(a2), it is found that the elements are divided into three distinct groups: $v_{3,i}\approx 0$ for $i=1,\ldots,30$, $v_{3,j}\approx -0.14$ for $j=31,\ldots,90$ and $v_{3,l}\approx 0.14$ for $l=61,\ldots,90$. The matrix of eigenvector distance for this case is plotted in Fig.~\ref{fig2}(c1), which shows that the nodes are divided into three distinct groups, $\mathbf{C}_1=(1,\ldots,30)$, $\mathbf{C}_3=(31,\ldots,60)$ and $\mathbf{C}_3=(61,\ldots,90)$, with $\delta e\approx 0$ for nodes within the same group but is larger than zero for nodes from different groups. The theory thus predicts that as $\varepsilon$ decreases from $\varepsilon_3$, the giant cluster formed in the previous stage [$\varepsilon\in(\varepsilon_3,\varepsilon_2)$] will be broken into two small-size clusters, and now there are three synchronization clusters on the network. Finally, as $\varepsilon$ crosses $\varepsilon_4$, the mode of $\lambda_4$ will be unstable. As depicted in Fig.~\ref{fig2}(a2), the elements of $\mathbf{v}_4$ are randomly distributed and, as a result of this, no eigenvector distance is close to $0$ and no clear structure is observed in the matrix [see Fig.~\ref{fig2}(d1)], indicating that the network is fully desynchronized in the case. 

The above predictions are well verified by numerical simulations. Setting $\varepsilon=7.6$, we plot in Fig.~\ref{fig2}(b2) the matrix of the synchronization error $\delta x_{i,j}=\left<|\mathbf{x}_i-\mathbf{x}_j|\right>_T$. We see that, in consistent with the theoretical prediction shown in Fig.~\ref{fig2}(b1), the oscillators are synchronized into two clusters, $\mathbf{C}_1=\mathcal{C}_1$ and $\mathbf{C}_2=\mathcal{C}_2\cup\mathcal{C}_3$. Plotted in Fig.~\ref{fig2}(c2) is the synchronization-error matrix for the coupling strength $\varepsilon=6$, which corresponds to the situation when the modes of $\lambda_2$ and $\lambda_3$ are unstable. We see that, in agreement with the predictions shown in Fig.~\ref{fig2}(c1), the oscillators are synchronized three clusters, $\mathbf{C}_1=\mathcal{C}_1$, $\mathbf{C}_2=\mathcal{C}_2$ and $\mathbf{C}_3=\mathcal{C}_3$. Finally, we plot in Fig.~\ref{fig2}(d2) the synchronization-error matrix for the coupling strength $\varepsilon=0.5$. In consistent with the predictions in Fig.~\ref{fig2}(d1), we see that there is no clear synchronization cluster and the network is fully desynchronized.

\begin{figure*}[tbp]
\begin{center}
\includegraphics[width=0.65\linewidth]{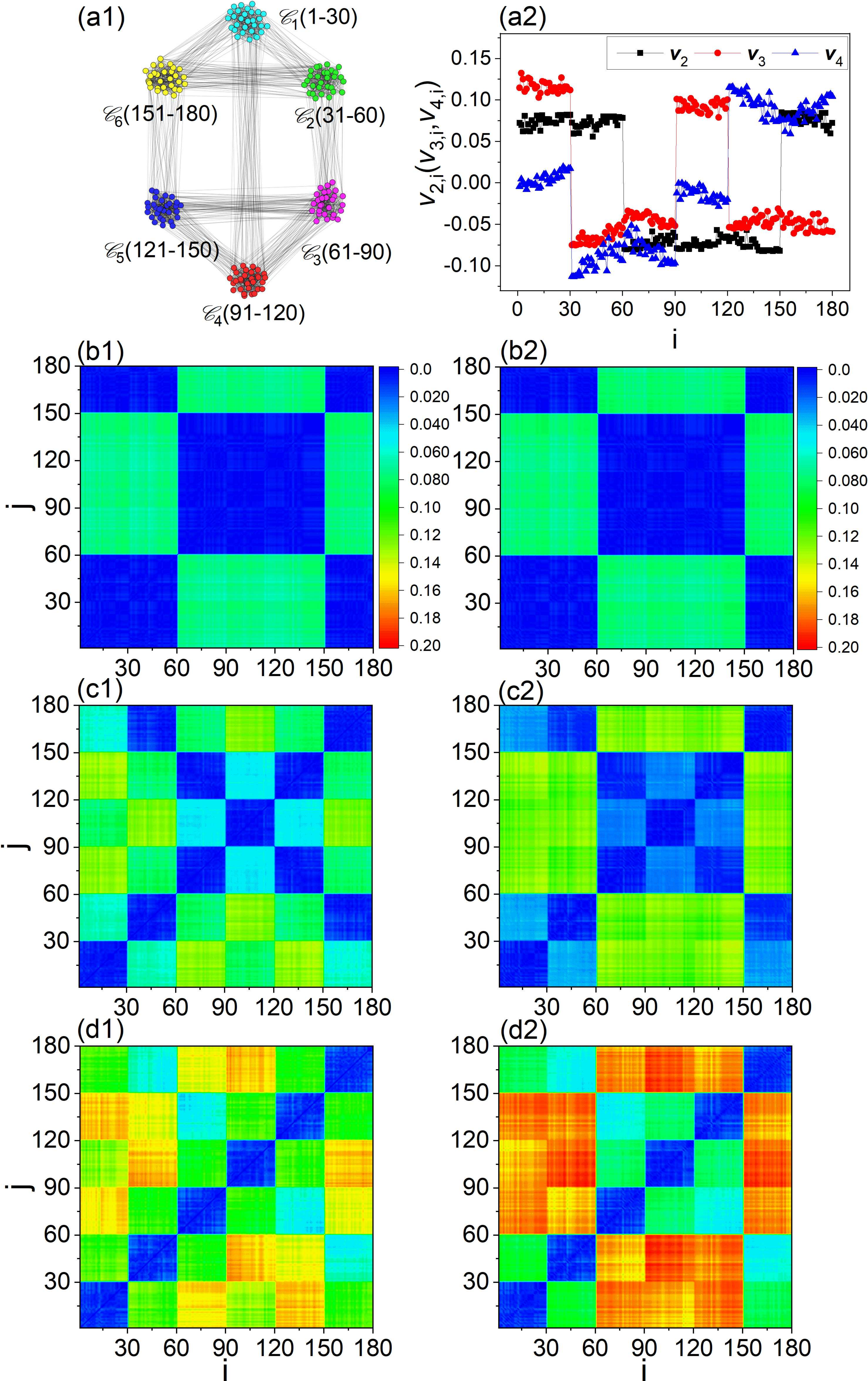}
\caption{(a1) The structure of the community network consisting of $N=180$ nodes and $M=6$ communities. (a2) The elements of the eigenvectors $\mathbf{v}_2$, $\mathbf{v}_3$ and $\mathbf{v}_4$. (b1) The matrix of eigenvector distance when the mode of $\lambda_2$ is unstable. (c1) The matrix of eigenvector distance when the modes of $\lambda_2$ and $\lambda_3$ are unstable. (d1) The matrix of eigenvector distance when the modes of $\lambda_2$, $\lambda_3$ and $\lambda_4$ are unstable. (b2-d2) are the matrices of synchronization error, $\{\delta x_{i,j}\}_{i,j=1,\ldots,N}$, obtained by numerical simulations for the coupling strengths $\varepsilon=10$ (b2), $\varepsilon=5.65$ (c2) and $\varepsilon=4$ (d2).}
\label{fig3}
\end{center}
\end{figure*}

We next investigate the CS behaviors in a large-size community network of hierarchical structures. The network structure is plotted in Fig.~\ref{fig3}(a1), which consists of $M=6$ communities and each community contains $n=30$ nodes. The network is generated by replacing each node in the 6-node network plotted in Fig.~\ref{fig1}(a) with a community. That is, the inter-connections are established only between the paired communities. To capture the weighted feature of the 6-node network, the inter-connecting probability between communities $2$ and $6$ (and also between communities $3$ and $5$) is set as $p_2=0.15$, while the inter-connecting probability for other paired communities is set as $p_2=0.1$. The connecting probability between nodes within each community is set uniformly as $p_1=0.9$. Still, the couplings are normalized and the nodal dynamics is described by the chaotic HR oscillator. The network nodes are ordered according to the partition of the communities, $\mathcal{C}_m=((m-1)n+1,\ldots,mn)$, with $m=1\ldots,M$ the community index. The three non-trivial largest eigenvalues of the network coupling matrix are $\lambda_2=-8.2\times 10^{-2}$, $\lambda_3=-0.136$ and $\lambda_4=-0.167$, and the corresponding critical couplings are $\varepsilon_2\approx 11.46$, $\varepsilon_3\approx 6.9$ and $\varepsilon_4\approx 5.6$. Shown in Fig.~\ref{fig3}(a2) are the elements of the eigenvectors $\mathbf{v}_{2}$, $\mathbf{v}_{3}$ and $\mathbf{v}_{4}$. As $\varepsilon$ decreases and crosses $\varepsilon_2$, the mode of $\lambda_2$ will be unstable. A check of the elements of $\mathbf{v}_2$ in Fig.~\ref{fig3}(a2) shows that the nodes are divided into two groups, $\mathbf{C}_1=\mathcal{C}_1\cup \mathcal{C}_2\cup \mathcal{C}_6$ (in which $v_{2,i}\approx 0.075$) and $\mathbf{C}_2=\mathcal{C}_3\cup \mathcal{C}_4\cup \mathcal{C}_5$ (in which $v_{2,i}\approx -0.075$). The corresponding matrix of eigenvector distance is plotted in Fig.~\ref{fig3}(b1), which, according to the framework of eigenvector-based analysis, implies that the oscillators will be synchronized into two clusters, $\mathbf{C}_1$ and $\mathbf{C}_2$, when the mode of $\lambda_2$ is destabilized. As $\varepsilon$ crosses $\varepsilon_3$, the mode of $\lambda_3$ will be also unstable. By checking the elements of both $\mathbf{v}_2$ and $\mathbf{v}_3$ in Fig.~\ref{fig3}(a2), we see that the nodes are divided into four distinct groups, $\mathbf{C}_1=\mathcal{C}_1$, $\mathbf{C}_2=\mathcal{C}_2\cup\mathcal{C}_6$, $\mathbf{C}_3=\mathcal{C}_4$ and $\mathbf{C}_4=\mathcal{C}_3\cup\mathcal{C}_5$. The matrix of eigenvector distance is plotted in Fig.~\ref{fig3}(c1), which suggests that the oscillators will be synchronized into four clusters when the modes of $\lambda_2$ and $\lambda_3$ are unstable. Finally, when the mode of $\lambda_4$ is unstable (as $\varepsilon$ crosses $\varepsilon_4$), the elements of $\mathbf{v}_2$, $\mathbf{v}_3$ and $\mathbf{v}_4$ in Fig.~\ref{fig3}(a2) show that the nodes are divided into six distinct groups, $\mathbf{C}_m=\mathcal{C}_m$, with $m=1,\ldots,6$. The corresponding matrix of eigenvector distance is plotted in Fig.~\ref{fig3}(d1). 

The above analysis thus predicts the following scenario of network desynchronization. With the decrease of the coupling strength, the whole network is firstly separated into two larger synchronization clusters (at $\varepsilon_2$), with one cluster formed by communities $1$, $2$ and $6$ and the other one formed by communities $3$, $4$ and $5$. Then, as $\varepsilon$ cross $\varepsilon_3$, each cluster is broken into two small-size clusters, and there are in total $4$ synchronization clusters on the network. In this stage, communities $2$ and $6$ are still synchronized, and also communities $3$ and $5$. Finally, as $\varepsilon$ falls below $\varepsilon_4$, community $2$ ($3$) is desynchronized from community $6$ ($5$), and there are $6$ synchronization clusters on the network in total, with each cluster corresponding to one community. These predictions are well verified by numerical simulations, as depicted in Fig.~\ref{fig3}(b2) [corresponding to the prediction in Fig.~\ref{fig3}(b1)], Fig.~\ref{fig3}(c2) [corresponding to the prediction in Fig.~\ref{fig3}(c1)] and Fig.~\ref{fig3}(d2) [corresponding to the prediction in Fig.~\ref{fig3}(d1)]. 

\subsection{Empirical neural networks}

\begin{figure}[tbp]
\begin{center}
\includegraphics[width=0.8\linewidth]{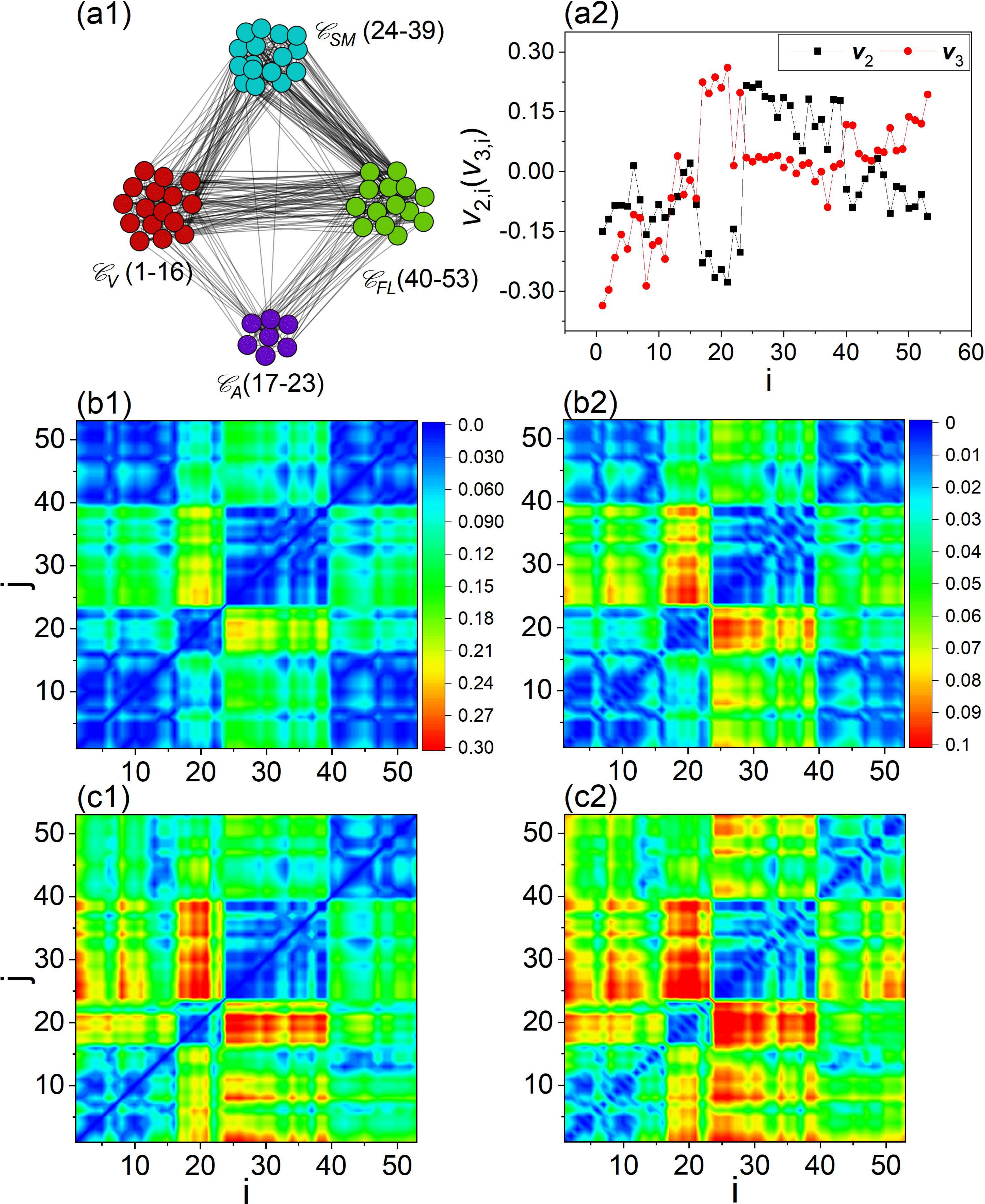}
\caption{CS in the cortico-cortical network of the cat brain. (a1) The network structure. The nodes are divided into four functional divisions: $\mathcal{C}_V=(1,\ldots,16)$ (visual division), $\mathcal{C}_A=(17,\ldots,23)$ (auditory division), $\mathcal{C}_{SM}=(24,\ldots,39)$ (somatomotor division) and $\mathcal{C}_{FL}=(40,\ldots,53)$ (frontolimbic division). (a2) The components of the eigenvectors $\mathbf{v}_{2}$ and $\mathbf{v}_{3}$. (b1) The matrix of eigenvector distance, $\{\delta e_{i,j}\}_{i,j=1,\ldots,N}$, when the mode of $\lambda_2$ is unstable. (c1) The matrix of eigenvector distance when the modes of $\lambda_2$ and $\lambda_3$ are unstable. (b2) The matrix of synchronization error, $\{\delta x_{i,j}\}_{i,j=1,\ldots,N}$, for the coupling strength $\varepsilon=2.2$. (c2) The matrix of synchronization error for $\varepsilon=1.9$.}
\label{fig4}
\end{center}
\end{figure}

We finally employ the framework of eigenvector-based analysis to investigate the CS behaviors in two empirical neural networks. The first example is the cortical network of the cat brain~\cite{CSinbN:Zhou1,CSCN:Scannell}.  The network structure is plotted in Fig.~\ref{fig4}(a1), which are constructed by $N=53$ nodes (cortex areas) and $L=830$ links (fiber connections). According to their functions, the cortex areas are divided into four divisions of variant size: $16$ areas in the visual division [$\mathcal{C}_V=(1,\ldots,16)$], $7$ areas in the auditory division [$\mathcal{C}_A=(17,\ldots,23)$], $16$ areas in the somatomotor division [$\mathcal{C}_{SM}=(24,\ldots,39)$], and $14$ areas in the frontolimbic division [$\mathcal{C}_{FL}=(40,\ldots,53)$]. (We note that the order of the neurons does not affect the synchronization behaviors of the network.) Still, the strategy of normalized couplings is adopted in constructing the network coupling matrix, and the chaotic HR oscillator is adopted to describe the nodal dynamics. The two largest non-trivial eigenvalues of the network coupling matrix are $\lambda_2=-0.395$ and $\lambda_3=-0.432$. The critical couplings of the two modes are $\varepsilon_2\approx 2.38$ (which is also the critical coupling for global synchronization) and $\varepsilon_3\approx 2.18$. The elements of the eigenvenctors $\mathbf{v}_2$ and $\mathbf{v}_3$ are plotted in Fig.~\ref{fig4}(a2). We see that, compared with the eigenvectors of the artificial networks [e.g., the results in Figs.~\ref{fig2}(a2) and \ref{fig3}(a2)], the elements of $\mathbf{v}_2$ and $\mathbf{v}_3$ are not clearly grouped. 

Shown in Fig.~\ref{fig4}(b1) is the matrix of eigenvector distance calculated for the coupling strength $\varepsilon=2.2$, by which only the mode of $\lambda_2$ is unstable. We see that the nodes are partitioned into three distinct clusters, $\mathbf{C}_1=\mathcal{C}_{V}\cup\mathcal{C}_{FL}$, $\mathbf{C}_2=\mathcal{C}_{A}$ and $\mathbf{C}_3=\mathcal{C}_{SM}$. Shown in Fig.~\ref{fig4}(c1) is the matrix of eigenvector distance for $\varepsilon=1.95$, by which both the modes of $\lambda_2$ and $\lambda_3$ are unstable. We see that in this case, in agreement with the neural divisions, the nodes are partitioned into four clusters, $\mathbf{C}_1=\mathcal{C}_{V}$, $\mathbf{C}_2=\mathcal{C}_{A}$, $\mathbf{C}_3=\mathcal{C}_{SM}$ and $\mathbf{C}_4=\mathcal{C}_{FL}$. To check the accuracy of the predictions, we plot in Fig.~\ref{fig4}(b2) and Fig.~\ref{fig4}(c2) the matrix of synchronization error, $\{\delta x_{i,j}\}_{i,j=1,\ldots,N}$, for the coupling strengths $\varepsilon=2.2$ and $1.9$, respectively. We see that the synchronization patterns predicted by the framework of eigenvector-based analysis [Figs.~\ref{fig4}(b1) and (c1)] are in good agreement with the ones obtained by model simulations [Figs.~\ref{fig4}(b2) and (c2)].

\begin{figure*}[tbp]
\begin{center}
\includegraphics[width=0.9\linewidth]{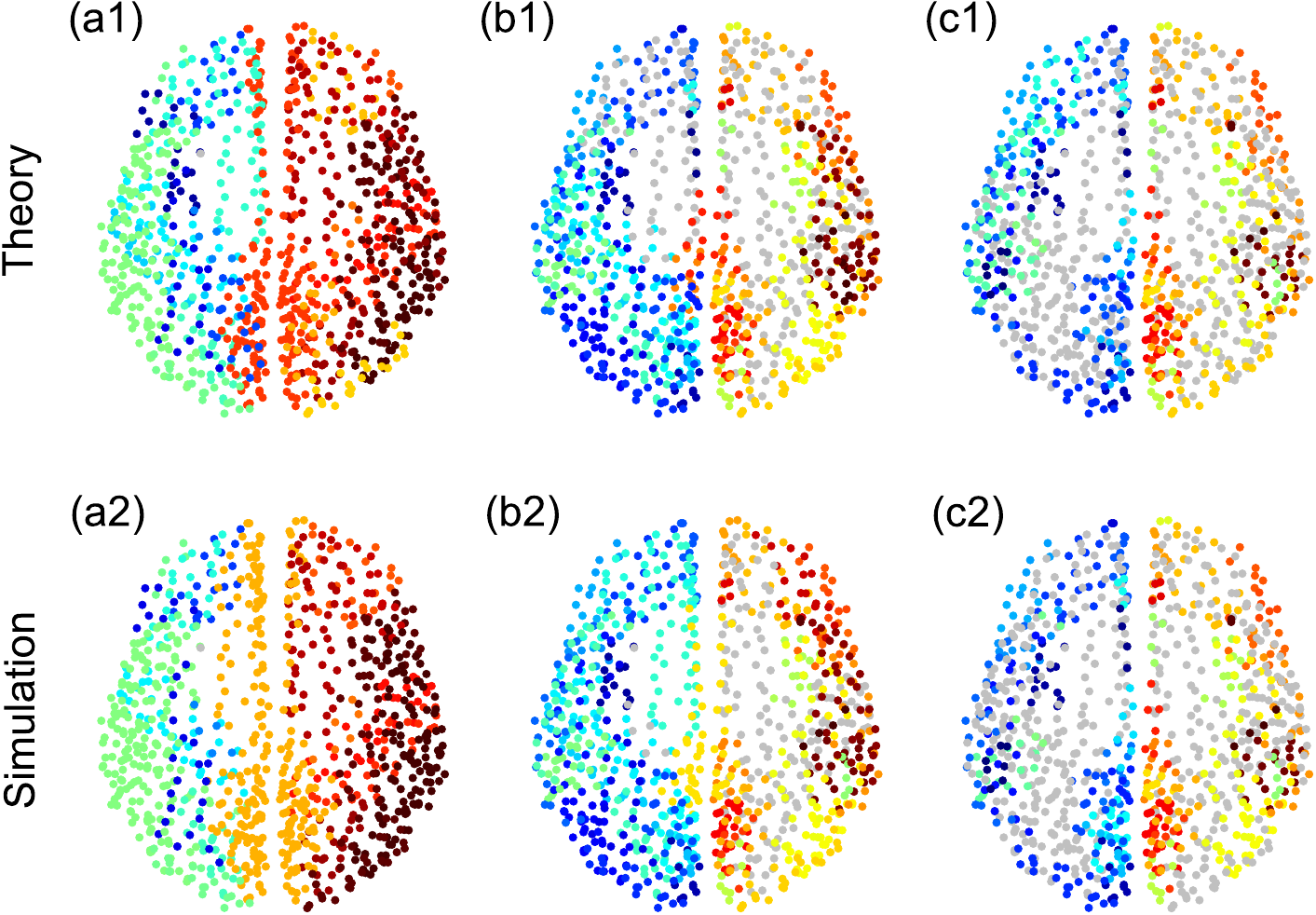}
\caption{CS in the cortical network of the human brain. The synchronization patterns predicted by the theory under different coupling strengths are plotted in the first row, and the corresponding results obtained by model simulations are plotted in the second row. (a1,a2) show the results for $\varepsilon=15$, by which only the mode of $\lambda_2$ is unstable. (b1,b2) are the results for $\varepsilon=8.5$, by which the modes of $\lambda_{2,3,4}$ are unstable. (c1,c2) are the results for $\varepsilon=5.5$, by which the modes $\lambda_{2,\ldots,6}$ are unstable. In each subplot, nodes in the non-synchronizable regions are represented by grey symbols, and nodes in the synchronizable regions are represented by colored symbols. Nodes with the same color (except the grey-colored nodes) form a synchronization cluster. See the context for more details.}
\label{fig5}
\end{center}
\end{figure*}

The second empirical neural network we consider is the cerebral cortex of the human brain~\cite{CSCN:Huo,CSCN:Hagmann,CSCN:Honey}, which contains $N=989$ nodes ($496$ nodes in the right hemispheres and $493$ nodes in the left hemispheres) and $L=17865$ links. According to the cytoarchitecture and functional parcellation, the nodes are partitioned into $M=64$ cortical regions (communities). Still, we adopt the normalized coupling strategy and use the chaotic HR oscillator to describe the nodal dynamics. The leading non-trivial eigenvalues of the network coupling matrix are $(\lambda_2,\lambda_3,\lambda_4,\lambda_5,\lambda_6)=(-3.7\times 10^{-2},-7.7\times 10^{-2}, -0.1, -0.12, -0.16, -0.18)$, and the corresponding critical couplings are about $(\varepsilon_2,\varepsilon_3,\varepsilon_4,\varepsilon_5,\varepsilon_6)=(25,12, 9.4, 8.3, 5.9, 5.2)$. Plotted in Fig.~\ref{fig5}(a1) are the synchronization patterns predicted by the framework of eigenvector-based analysis for the coupling strength $\varepsilon=15$ (by which only the mode of $\lambda_2$ is unstable). In plotting Fig.~\ref{fig5}(a1), we first calculate for each cortical region the averaged eigenvector distance $\left<\delta e\right>_m=\sum_{i,j\in{\mathcal{C}_m}}\delta e_{i,j}/n_m(n_m-1)$, with $\mathcal{C}_m$ denoting the set of nodes in the $m$th region, $n_m$ is the size of the $m$th region, and $\delta e_{i,j}$ is the eigenvector distance between nodes $i$ and $j$. The nodes within the $m$th region are regarded as synchronizable if $\left<\delta e\right>_m$ is smaller than the threshold $\left<\delta e\right>_c=0.01$. By doing so, the $M=64$ regions are divided into two groups, the synchronizable and non-synchronizable groups. The nodes in the non-synchronizable regions are shown in grey, which are excluded from the further analysis. Having identified the set of synchronizable regions, we next partition the regions into clusters based on the eigenvector distance between them. Here the eigenvector distance between regions $m$ and $m'$ is defined as $\left<\delta e\right>_{m,m'}=\sum_{i\in\mathcal{C}_m; j\in\mathcal{C}_{m'}}\delta e_{i,j}/n_m n_{m'}$, with $\mathcal{C}_m$ and $\mathcal{C}_{m'}$ denoting the sets of nodes in the $m$th and $m'$th regions, respectively, and $\delta e_{i,j}$ is the eigenvector distance betwteen nodes $i$ and $j$. Finally, regions with $\left<\delta e\right>_{m,m'}<0.003$ are considered as synchronizable and nodes in these regions are marked by the same color. We see in Fig.~\ref{fig5}(a1) that most of the nodes are colored, indicating a higher synchronization degree of the whole network. Meanwhile, the colored nodes are organized into clusters of different sizes and specific spatial distributions. By the same procedure, we plotted in Figs.~\ref{fig5}(b1) and (c1) the synchronization patterns predicted by the theoretical framework for the coupling strengths $\varepsilon=8.5$ (by which the modes of $\lambda_2$, $\lambda_3$ and $\lambda_4$ are unstable) and $\varepsilon=5.5$ (by which the modes of $\lambda_{2,\ldots,6}$ are unstable). We see that, as more modes become unstable, more nodes are shown in grey, indicating the degraded synchronization performance of the whole network. In the meantime, with the decrease of the coupling strength, the sizes of the clusters are shrunk gradually.

The above predictions are well verified by simulations. Setting $\varepsilon=15$, we plot in Fig.~\ref{fig5}(a2) the synchronization pattern obtained by model simulations. In plotting Fig.~\ref{fig5}(a2), we first calculate the node-averaged synchronization error of each region, $\left<\delta x\right>_m=\sum_{i,j\in{\mathcal{C}_m}}\delta x_{i,j}/n_m(n_m-1)$, with $\delta x_{i,j}=\left<|x_i-x_j|\right>_T$ is the time-averaged synchronization error between nodes $i$ and $j$. The nodes within the $m$th region are regarded as synchronizable if $\left<\delta x\right>_m$ is smaller than the threshold value $\left<\delta x\right>_c=0.035$. Still, nodes in the non-synchronizable regions are shown in grey. We then partition the synchronizable regions into clusters based on the synchronization error $\left<\delta x\right>_{m,m'}=\sum_{i\in\mathcal{C}_m; j\in\mathcal{C}_{m'}}\delta x_{i,j}/n_m n_{m'}$. The $m$th and $m'$th regions are regarded as synchronized if $\left<\delta x\right>_{m,m'}<0.01$, and nodes in the synchronized regions are marked by the same color. We see that the synchronization pattern obtained by simulations [Fig.~\ref{fig5}(a2)] is in good agreement with the pattern predicted by the theory [Fig.~\ref{fig5}(a1)]. The good agreement between simulation and prediction is also observed for the coupling strengths $\varepsilon=8.5$ and $\varepsilon=5.5$, as depicted in Fig.~\ref{fig5}(b2) [corresponding to the prediction in Fig.~\ref{fig5}(b1)] and Fig.~\ref{fig5}(c2) [corresponding to the prediction in Fig.~\ref{fig5}(c1)]. We note that by changing the thresholds $\left<\delta e\right>_c$ and $\left<\delta x\right>_c$, the sizes of the clusters will be changed, but the consistency between the theoretically predicted and numerically obtained patterns is kept unchanged, which has been checked by additional simulations (not shown).     

\section{Discussions and conclusion}

Whereas the significant impacts of network structure on synchronization have been well recognized and demonstrated in the literature, attention has been mainly focused on the roles of the extreme modes of the network coupling matrix. Specifically, for complex networks of linearly coupled identical chaotic oscillators, the MSF formalism suggests that the synchronizability of the networks can be well characterized by the extreme eigenvalues of the network coupling matrix~\cite{MSF-1}, i.e., the largest non-trivial eigenvalue ($\lambda_2$) or the ratio between the largest and smallest non-trivial eigenvalues ($\rho=\lambda_N/\lambda_2$). For this reason, a large body of studies on network synchronization has been focusing on the dependence of the extreme eigenvalues on the network structure or the coupling strategy~\cite{REV:SB2006,REV:Arenas}, with the impacts and roles of the other modes being largely overlooked. While analysis based on extreme eigenvalues is efficient and convenient in applications, it applies to only the special case of global synchronization state but not the CS states appeared in the transition to global synchronization. Our studies in the present work show that, to fully characterize the synchronization behaviors on complex networks and the CS states in synchronization transition in particular, it is equally important to consider the impacts and roles of the modes with moderate eigenvalues.

Our present work also highlights the significance of eigenvectors in exploring the CS behaviors of complex networks. Previously, eigenvectors of the network coupling matrix have been exploited to investigate the synchronization dynamics of networked systems, but the studies are limited to slightly desynchronized networks or small-size network motifs~\cite{CSev:Fu,CSev:Poel,CSev:Khanra}. In Ref.~\cite{CSev:Fu}, the authors studied the synchronization behavior of chaotic oscillators in large-size complex networks at the boundary of global synchronization, and found that the stabilities of the oscillators can be well predicted from the elements of the eigenvector associated with the unstable mode. In Ref.~\cite{CSev:Poel}, the authors investigated the dynamical patterns on small-size networks of up to $5$ nodes, and found that, when the eigenvectors of the network coupling matrix satisfy some special properties, the dynamical patterns can be well inferred from the network topology and the local dynamics. In Ref.~\cite{CSev:Khanra}, the authors demonstrated that, by the information of the eigenvector of the leading mode, all synchronization clusters in a complex network can be identified without knowing the network symmetries and, in addition, their sequence in the process of synchronization transition can be predicted. Compared to the existing studies, the framework proposed in our current study is featured by the capability of analyzing the synchronization patterns (CS states) formed in deeply desynchronized networks (with the number of unstable modes unlimited) and, more importantly, in large-size complex networks of non-perfect symmetries. The latter makes the two different approaches currently employed in CS studies, namely the symmetry-based (in which perfect network symmetry is required and the clusters are defined on complete synchronization)~\cite{ADM:2016,CS:BAO,CS:OTT2007,SyncPattern,CS:WXG2014,SynPat:Schaub,FS:2016,JDH:2019,Pecora2014,Recentadvances,LWJ-1,LWJ-2,NTTSCS,YC:2017,BC:2018,CS:WYF2019,CSWL:2020} and community-based approaches (in which no perfect network symmetry is required and the synchronization clusters are loosely defined)~\cite{CSinbN:Zhou1,CSinbN:Zhou2,CSCN:Huo,commnet:Huang,commnet:Wang}, unified within the same framework, thereby paving a way to the exploration of the synchronization patterns in real-world systems.  

The key result we have obtained in the theoretical analysis is given by Eq.~(\ref{theory}), which reveals how the synchronization patterns are connected to the eigenvectors of the network coupling matrix. In obtaining the Eq.~(\ref{theory}), we have made the assumption that the statistical properties of the coupled oscillators in CS are close to that of the isolated oscillator. The purpose of this assumption (approximation) is to make the same master equation, i.e. the Eq.~(\ref{msf}), applicable to all the perturbation modes, so that the critical couplings where the modes become unstable can be analytically estimated ($\varepsilon_i=-\sigma_c/\lambda_i$). While the validity of this approximation has been verified by numerical simulations in our current work, there are cases where the approximation is invalid~\cite{WY:2022}. In these invalid cases, we are still able to identify the synchronization patterns (the CS states) according to the eigenvector matrix, but are unable to predict analytically the critical couplings where the patterns are generated. (In cases like this, the stability of each CS state should be evaluated individually, e.g., by calculating the largest conditional Lyapunov exponent~\cite{CS:WYF2019,Pecora2014}.) 

An alternative approach to analyzing the CS behaviors in networked chaotic oscillators could be the energy-balance-based method~\cite{MJ:2023}. Like the method of Lyapunov function, the CS states stand as the local minima of the energy landscape. Our study suggests that these local minima are determined by the network symmetries and can be inferred from the eigenvectors of the network coupling matrix. A systematic analysis of the application of the energy-balance-based method to CS behaviors is out of the scope of our current study.

To summarize, exploiting the eigenvectors of the network coupling matrix, we have proposed a new theoretical framework to explore the CS behaviors in general complex networks. The new framework, which requires no prior knowledge of the network symmetries, is able to predict not only all the CS states generated in the process of synchronization transition, but also the critical couplings where the CS states are generated and the sequence of the CS states in the transition. The efficacy and efficiency of the framework have been verified by a variety of network models of identical chaotic oscillators, including small-size artificial networks of perfect symmetries, large-size complex networks of distinct community structures, and large-size empirical neural networks. Our studies highlight the importance of eigenvectors in exploring CS in complex networks, and the framework we have proposed provides a powerful tool to the exploration of synchronization patterns in real-world complex systems.

\section*{Acknowledgement}
This work was supported by the National Natural Science Foundation of China (NSFC) under Grant Nos. 12105165 and 12275165. XGW was also supported by the Fundamental Research Funds for the Central Universities under Grant No. GK202202003.

\end{document}